\documentclass[a4paper]{spie}  
\usepackage[]{graphicx}
\usepackage{amssymb,amstext}
\usepackage{cite}

\def\sun{\hbox{$\odot$}}

\def\arcsec{\hbox{$^{\prime\prime}$}}

\title{New opportunities with spectro-interferometry and spectro-astrometry}

\author{Stefan Kraus
\skiplinehalf
Department of Astronomy, University of Michigan, 500 Church St., Ann Arbor, MI 48109, USA
}

\authorinfo{Further author information:\\
E-mail: stefankr@umich.edu, Telephone: +1 734 615 7374}

\begin{document} 
  \maketitle 

\begin{abstract}
Latest-generation spectro-interferometric instruments combine a
milliarcsecond angular resolution with spectral capabilities, 
resulting in an immensely increased information content.
Here, I present methodological work and results that 
illustrate the fundamentally new scientific insights provided 
by spectro-interferometry with very high spectral dispersion 
or in multiple line transitions (Brackett and Pfund lines).
In addition, I discuss some pitfalls 
in the interpretation of spectro-interferometric data.
In the context of our recent studies on the classical Be stars
$\beta$~CMi and $\zeta$~Tau, I present the first
position-velocity diagram obtained with optical interferometry 
and provide a physical interpretation for a phase inversion,
which has in the meantime been observed for several classical Be-stars.
In the course of our study on the Herbig B[e] star V921~Sco,
we combined, for the first time, spectro-interferometry and spectro-astrometry, 
providing a powerful and resource-efficient way to constrain the 
spatial distribution as well as the kinematics of the circumstellar
gas with an unprecedented velocity resolution up to $R=\lambda/\Delta\lambda=100,000$.
Finally, I discuss our phase sign calibration procedure,
which has allowed us to calibrate AMBER differential phases and 
closure phases for all spectral modes, and derive from the gained
experience science-driven requirements for future instrumentation
projects.
\end{abstract}



\section{INTRODUCTION}
\label{sec:intro}

One of the most remarkable features of the latest-generation optical 
interferometric instruments is their capability to combine 
a milliarcsecond angular resolution with high spectral dispersion,
enabling velocity-resolved studies in spectral lines.
Spectro-interferometric instruments with sufficient resolution for
velocity-resolved studies ($R\gtrsim 1000$) have now been implemented
at various facilities operating at near-infrared 
({VLTI/AMBER\cite{pet07}}: up to $R=12,000$; 
{Keck/V2-SPR\cite{woi10}}: $R=1700$),
mid-infrared ({ISI\cite{mon00}}: up to $R\sim 100,000$), and 
visual wavelengths ({CHARA/VEGA\cite{mou08}}: up to $R=35,000$).
The primary observables provided by these instruments are spectra
and wavelength-differential visibilities, phases (DPs), and
closure phases (CPs, in the case of the combination of 3 or more
apertures).

The aim of this paper is to discuss some applications enabled by this
exciting new technique, to outline challenges and
pitfalls in the astrophysical interpretation, and to identify the most
important limiting factors of this technique. 
In addition, I would like to highlight the highly complementary
nature of optical long-baseline spectro-interferometry and 
spectro-astrometric observations obtained with
conventional high-resolution spectrographs.
For this purpose, I will present studies that we have conducted on the
classical Be stars {$\beta$~CMi\cite{kra12a}} and
{$\zeta$~Tau\cite{kra12a}} and the B[e] star
{V921\,Sco\cite{kra12b,kra12c}} using the AMBER instrument. 
Utilizing the VLTI 8.2m unit telescopes and the
fringe phase-tracking instrument {FINITO\cite{gai04,leb08}}, AMBER
covers an interesting parameter space in terms of high spectral dispersion
(enabling velocity-resolved studies), a good sensitivity (enabling
studies on fainter object classes), and the availability of multiple
baselines (enabling detailed studies of the 2-D gas distribution and
velocity field).
However, most of the presented strategies and arguments are not limited
to VLTI, but are applicable to spectro-interferometric instruments in
general.

\section{SPECTRO-INTERFEROMETRY}
\label{sec:spectrointerferometry}

\subsection{PHOTOCENTER ANALYSIS}
\label{sec:photocenter}

The {DP\cite{pet86,pet88}} information measured by spectro-interferometers provides a
particularly powerful observable to constrain the gas velocity
field in astrophysical objects. 
In first-order approximation, DPs measure the
displacement of the line emission with respect to the
continuum emission, although one has to be aware that this
approximation breaks down if the line-emitting region is significantly
resolved by the interferometer. 

Given that each baseline measures only the projection of the
{photocenter\cite{bec82}} displacement vector along the baseline vector, it is
highly desirable to obtain multiple measurements and to probe a wide
range of position angles (PAs). 
Such multi-baseline interferometric measurements allow one
to recover the 2-D photocenter displacement vector\cite{leb09}
$\vec{p}$ using the relation
\begin{equation}
  \vec{p} = -\frac{\phi_{i}}{2\pi} \cdot \frac{\lambda}{\vec{B_{i}}},
  \label{eq:photocenter}
\end{equation}
where $\phi_{i}$ is the DP measured on baseline $i$,
$\vec{B_{i}}$ is the corresponding baseline vector, 
and $\lambda$ is the central wavelength.
As noted above, this set of linear equations is only valid for
spatially unresolved or marginally resolved {objects\cite{pet89,lac03}}.
The equations are already overdetermined
with a single 3-telescope measurement ($i=3$), providing a useful
cross-check whether the assumption is justified or
whether the measured DPs trace higher-order geometric effects.
The precision of the derived photocenter vector depends both on the
achievable DP accuracy as well as the uniformity of the PA coverage
that has been achieved in the observations. 

\begin{figure}[p]
  \centering
  \includegraphics[width=10cm]{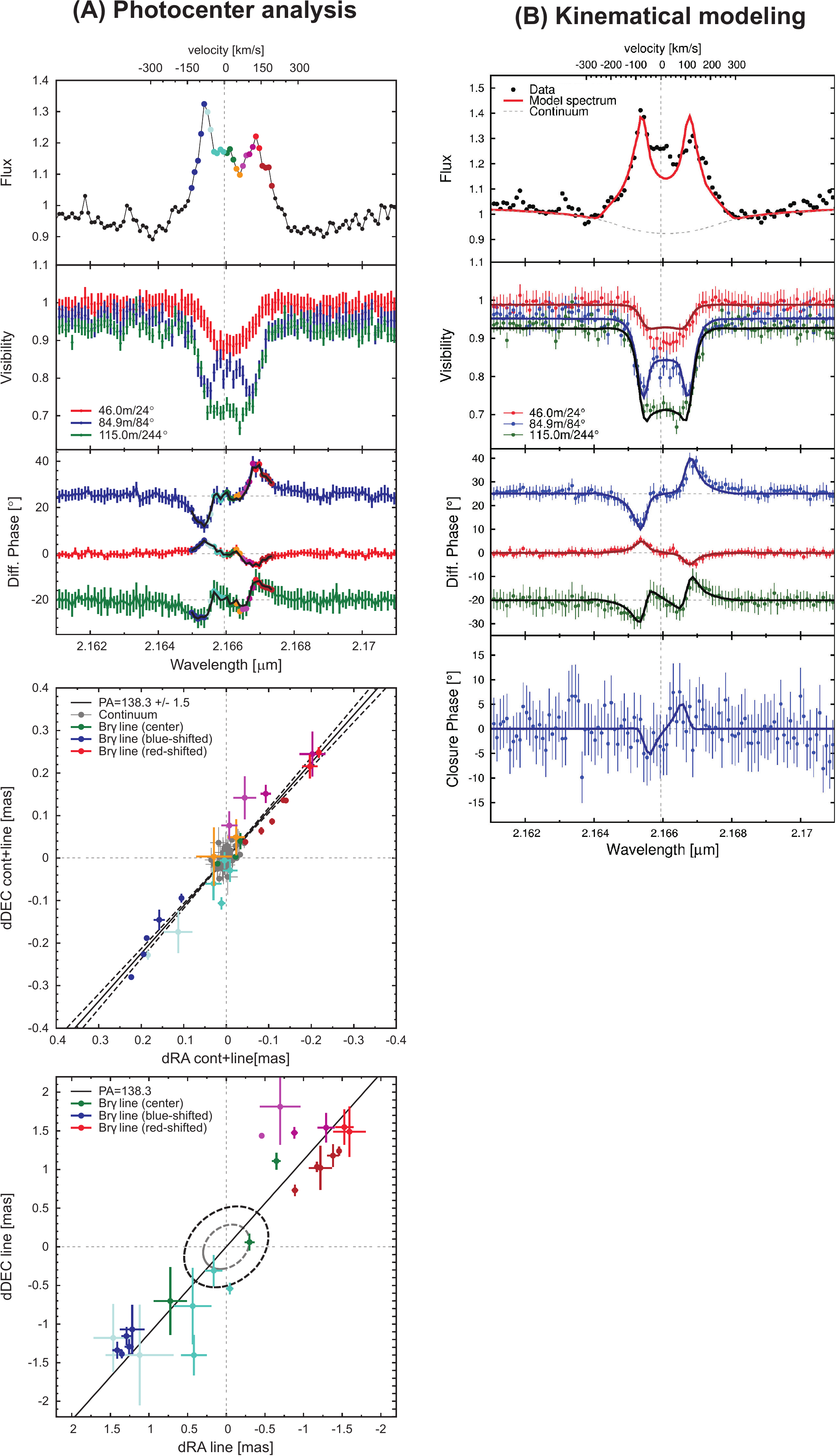}
  \caption{
    VLTI/AMBER spectro-interferometric observations
    obtained on the classical Be star {$\beta$~CMi\cite{kra12a}}.
    {\bf (A):} The top panel shows the measured line profile,
    visibilities, and DPs.
    Using eq.~\ref{eq:photocenter}, we derived the photocenter
    offsets (middle panel), which we then corrected for the
    continuum contributions using eq.~\ref{eq:photocentercorr2}
    (bottom panel; see Sect.~\ref{sec:photocenter} for details).
    {\bf (B):} Kinematical model that can reproduce the
    observed visibility and DP signatures, including a phase inversion
    in the line center, assuming a purely Keplerian velocity field
    (see Sect.~\ref{sec:modeling}).
  }
  \label{fig:BCMi}
\end{figure}

In case the photocenter analysis method is applicable, it can
provide basic, yet powerful information about the gas kinematics on
unprecedentedly small angular scales.
For instance, with a typical DP accuracy of $1^{\circ}$,
it is possible to measure photocenter displacements of
$\sim0.012$~milliarcseconds (mas), approaching the micro-arcsecond regime. 
In a recent {study\cite{kra12a}} we observed the classical
Be star $\beta$~CMi using AMBER's high spectral dispersion mode
($R=12,000$) and measured strong non-zero DPs, as well as a strong
visibility drop within the line (Fig.~\ref{fig:BCMi}A, top, left panel).
Applying eq.~\ref{eq:photocenter} to the data reveals that the blue-
and red-shifted line wings are displaced in opposite directions with
respect to the star (Fig.~\ref{fig:BCMi}A, middle, left panel).  The
PA of the line photocenter displacement ($138.3 \pm 1.5^{\circ}$) is
consistent with the measured orientation of the continuum-emitting 
circumstellar disk ($139.2^{+4.4\circ}_{-6.3}$, as determined with
CHARA/MIRC continuum interferometry\cite{kra12a}), indicating a
rotation-dominated velocity field.

Using the aforementioned procedure, it is possible to measure
the direction of the photocenter displacement, while the length of the
displacement vector is biased by the contributions from the
underlying continuum emission.   
In order to remove these contributions from the observables measured
by the interferometer ($F$, $V$, $\phi$), we interpolate the continuum
flux ($F_c$) and continuum visibility ($F_c$) from the adjacent continuum.
The visibility and DP of the pure line-emitting {region\cite{leb09}} ($V_l$, $\phi_l$) are then
given by
\begin{eqnarray}
  | F_l V_l |^2  & = &  | F V |^2 + | F_c V_c |^2 - 2 \cdot F V \cdot F_c V_c \cdot \cos \phi
  \label{eq:photocentercorr1}
\\
  \sin \phi_l   & = & \sin \phi \frac{| F V |}{| F_l V_l |},
  \label{eq:photocentercorr2}
\end{eqnarray}
where $F_l=F-F_c$ denotes the flux contribution from the spectral line.
Given that the quantities in eqs.~\ref{eq:photocentercorr1}+\ref{eq:photocentercorr2} are
associated with independent uncertainties, the continuum-corrected DP 
$\phi_l$ will exhibit a significantly larger uncertainty than the
measured continuum+line DP $\phi$, which also translates in larger
error bars for the derived line photocenter offset vectors (in
particular for weak spectral lines).

Applying this correction to our $\beta$~CMi data
(Fig.~\ref{fig:BCMi}A, bottom, left panel) yields that the
amplitude of the line displacement is about $\sim 1.5$~mas and that
the displacement in the blue- and red-shifted line wing is rather
symmetric, suggesting that $\beta$~CMi does not exhibit a significant
one-armed oscillation, as found for other Be {stars\cite{ste09}}.

It should be noted that eqs.~\ref{eq:photocentercorr1}+\ref{eq:photocentercorr2} require
absolute-calibrated visibilities, which are difficult to extract
from spectro-interferometric observations with high spectral dispersion.
In order to achieve a sufficient signal-to-noise ratio (SNR), such
observations require long detection integration times and an active
stabilization of the atmosphere-induced fringe phase modulations, 
in our study using the VLTI/FINITO instrument.
The fringe stabilization leaves residual phase jitter that degenerates
the visibility amplitude on the science beam combiner in 
an uncontrollable way.  As a result, the visibility amplitude cannot
be calibrated reliably, and complementary observations without fringe
tracking and in low spectral dispersion are required in order to
recalibrate the absolute visibility level. Effectively, this doubles
the total observing time and causes practical problems with respect to
matching the $uv$-sampling and to minimizing the time interval between the
two measurements, which constitutes a severe problem for time-variable
objects, such as Be-stars or young stellar objects.

\subsection{POSITION-VELOCITY DIAGRAMS}
\label{sec:posvel}

In radio interferometry, position-velocity diagrams constitute a
vital element for the analysis of kinematical signatures and
to constrain, for instance, the rotation profile of gas disks around
galactic {nuclei\cite{miy95}}.  
These diagrams can be constructed either from velocity-resolved
``channel maps'' or from astrometric measurements
in maser emission.

\begin{figure}[b]
  \centering
  \includegraphics[width=15cm]{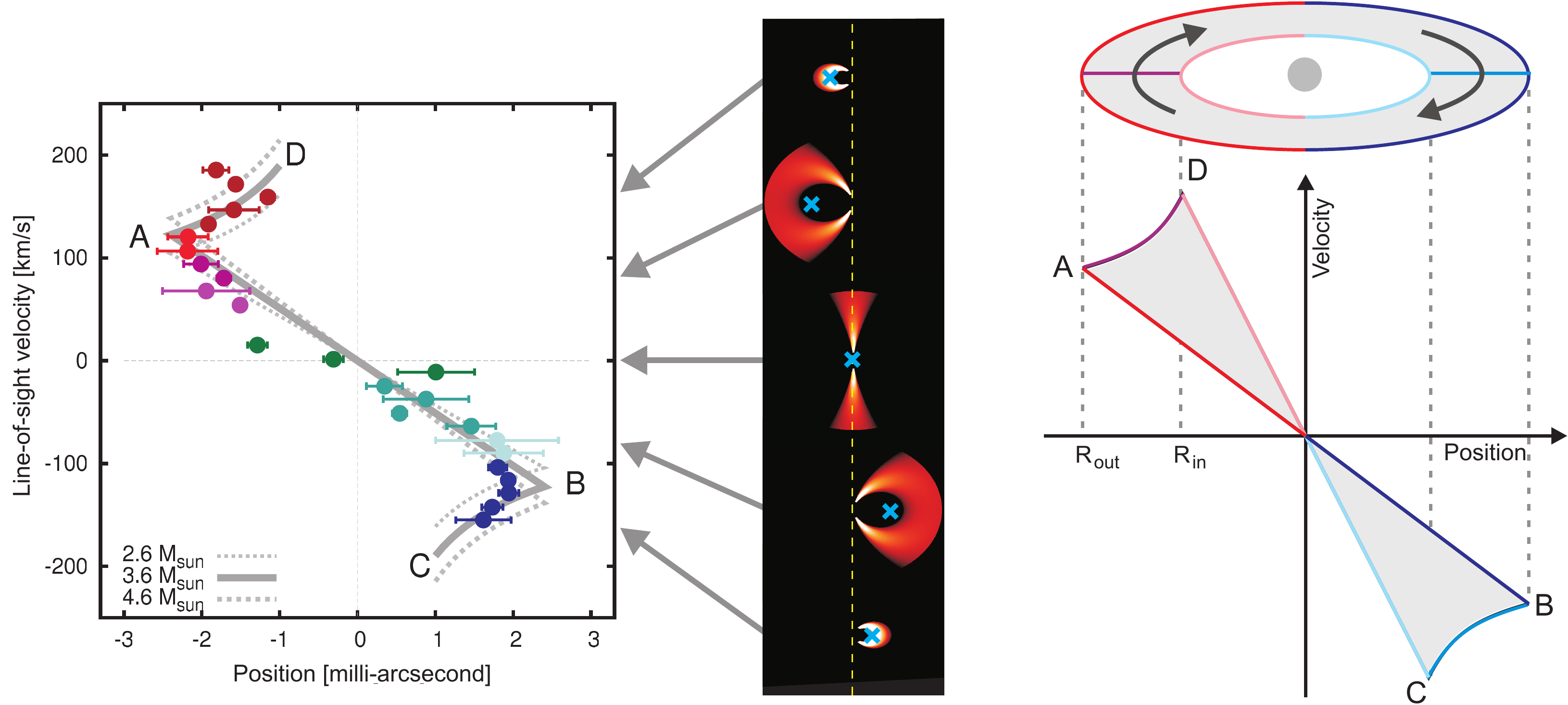}
  \caption{{\bf Left:} Position-velocity diagram for $\beta$~CMi, constructed
    from a single VLTI/AMBER observation with spectral dispersion
    $R=12,000$.
    {\bf Right:} Sketch, illustrating the ``bowtie''-shaped
    structures, which appear in the position-velocity diagrams of
    Keplerian-rotating disks (see Sect.~\ref{sec:posvel} for details).
  }
  \label{fig:posvel}
\end{figure}

In our study on {$\beta$~CMi\cite{kra12a}}, we constructed
an equivalent diagram based on the photocenter offsets derived from
our optical interferometry data.
For this purpose, we convert the wavelength of each spectral
channel to the corresponding Doppler velocity and plot it against the
length of the continuum-corrected photocenter vectors (which we
projected beforehand onto the determined disk plane;
Fig.~\ref{fig:BCMi}). 
The resulting diagram (Fig.~\ref{fig:posvel}, left) 
shows that the stellocentric emission radius of the line-emitting gas 
increases approximately linearly with velocity for low/medium
velocities  ($|v| \lesssim 100$~km\,s$^{-1}$; line A-B in
Fig.~\ref{fig:posvel}, left) and then reaches a maximum 
value at about $\sim 120$~km\,s$^{-1}$.  
At higher gas velocities ($120 \lesssim |v| \lesssim 200$~km\,s$^{-1}$; arcs A-D
and B-C in Fig.~\ref{fig:posvel}, left), the emission radius decreases again.
This behaviour can be explained in a Keplerian-rotation scenario,
assuming that the disk extends from an inner radius $R_{\mathrm in}$
to an outer radius $R_{\mathrm out}$ (Fig.~\ref{fig:posvel}, right):
If we first consider only gas orbiting at the outer radius $R_{\mathrm out}$, 
then it is clear that the lowest projected velocity appears along
the line-of-sight (position 0), while the maximum projected velocity
will appear at separation $R_{\mathrm out}$, as observed in line A-B.
Gas located at the inner radius $R_{\mathrm in}$ will occupy a
similar line in the position-velocity diagram, even though the maximum
separation will now be smaller ($R_{\mathrm in}$) and the maximum
velocity will be higher, given that the velocity in a
Keplerian-rotating disk increases with $v(r) \propto r^{-0.5}$.
Filling in the intermediate radii results then in the characteristic
``bowtie''-shaped region shown in Fig.~\ref{fig:posvel}
(right; grey filled area).
Comparing the size of the emitting area for the different emission
annuli, it is clear that our astrometric measurements are dominated
by the emission from the outer-most disk radius (line A-B) as well as the
high-velocity wings (arcs A-D and B-C), explaining the measured
position-velocity curve of $\beta$~CMi.
Based on these considerations, it is possible to construct a simple
analytic {model\cite{kra12a}}, which can be fitted to the derived position-velocity
diagram (Fig.~\ref{fig:posvel}, left; grey curve).  From this model,
we derive the mass of the central star to $3.5 \pm 0.2$~M$_{\sun}$,
which is in agreement with the value derived using a more
sophisticated modeling approach (Sect.~\ref{sec:modeling}).

Overall, position-velocity diagrams provide a very intuitive and
model-independent method for the interpretation of velocity-resolved
astrometric observations and can be constructed already from a single
3-telescope spectro-interferometric observation.

\subsection{KINEMATICAL MODELING: CONSTRAINING THE DETAILED
  GAS-VELOCITY FIELD}
\label{sec:modeling}

Photocenter offsets and position-velocity diagrams provide a
straightforward and model-independent method to obtain first-order
kinematical information, which can already be sufficient to discern
between competing models, such as rotating disks or the ejection of
material in bipolar outflows. In order to distinguish between
more complex velocity fields and to derive quantitative constraints, it
is necessary to employ a more sophisticated model fitting
approach.

Kinematical modeling was also crucial in solving a puzzle
concerning a phase inversion that appears in the center of the
Br$\gamma$-line in our $\beta$~CMi data.
Earlier spectro-interferometric observations on Be stars with lower
spectral dispersion (e.g.\ {$\alpha$~Arae\cite{mei07}}) have shown a 
simple S-shaped signature that can be interpreted very intuitively in
terms of a disk rotation scenario, where the approaching 
(blue-shifted) disk part is displaced in the opposite direction from
the receding (red-shifted) disk part.
However, our AMBER $R=12,000$ observations on $\beta$~CMi
(Fig.~\ref{fig:BCMi}) reveal a W-shaped profile, with a clear
phase sign inversion in the line center.
Stefl et {al.\cite{ste11}} observed this phase inversion also in
various other Be stars, including $\zeta$~Tau, $\alpha$~Col,
$\delta$~Sco, $\omega$~CMa, and 48~Lib, and proposed that the
phase inversion might indicate either secondary dynamical effects or
the need for an additional kinematical component beyond the canonical
star+disk paradigm, such as a polar jet.

In a first interpretation attempt, we investigated whether the phase
inversion might be explained with the presence of a hidden photospheric
Br$\gamma$ absorption component underlying the circumstellar
Br$\gamma$ emission. Assuming that the stellar photosphere rotates in the same
sense as the circumstellar disk, the DP signature induced by an
absorption line would cause a photocenter offset in the opposite
direction from a circumstellar emission line, which could
qualitatively explain the observed phase effect.
However, when modeling the expected DP signature using a stellar
atmosphere code for fast rotating {stars\cite{mon07,che11}}, we find
that these corresponding signatures ($\lesssim 0.5^{\circ}$) are 
by far too small to explain the measured phase signals of $\sim 15^{\circ}$.

In a second attempt, we considered whether the observed phase
inversion might be explained with the phase jump appearing at the
transition between different lobes of the visibility function.
In this scenario, a standard Keplerian disk model might already be
sufficient to explain the phase jumps, since the spatially most
extended structure would appear at the lowest gas velocities, i.e.\ in
the line center.  Accordingly, the spectral channels in the line wings
(corresponding to high gas velocities) would measure visibilities in
the first lobe of the visibility function, while the channels in the
line center (corresponding to low gas velocities) would trace the
second lobe.  This scenario is supported by the fact that the spectral
channels, where the DP phase jumps occur, also exhibit minima in the
measured wavelength-differential visiblity, tracing the transition
from the first to the second lobe.  

In order to check these qualitative arguments, we adopted our
kinematical modeling {code\cite{wei07,kra12a,kra12c}} to
simulate the gas velocity field of rotating disks with arbitrary
rotation profile $|v(r)| \propto r^{\beta}$, where $\beta=-0.5$
corresponds to Keplerian rotation. Our modeling approach assumes that
the line emission is located in a geometrically thin disk that
extends from the star to an outer radius $R_{\rm out}$.
The code computes synthetic channel maps at spectral resolution
$R=100,000$, which we convolve to the spectral resolution of
AMBER.  From these convolved images we compute spectra, visibilities,
DPs, and CPs for comparison with our data.
Besides the line emission, our model includes the continuum 
emission from the circumstellar disk and the photospheric emission of
a star rotating at near-critical velocity. 
Applying this continuum+line model to $\beta$~CMi we find that a
Keplerian-rotating disk indeed reproduces the observed phase
inversion also quantitatively. Varying both the geometric parameters
and the disk rotation profile, we find that we can obtain an excellent
fit to all observables assuming $\beta=-0.5 \pm 0.1$, with a mass of
the central object of $3.5 \pm 0.2$~M$_{\sun}$ (Fig.~\ref{fig:BCMi}B).
We conclude that the disk around $\beta$~CMi exhibits, to high
precision, a Keplerian rotation profile, which supports viscous
decretion disk models, where the Keplerian-rotating disk is 
replenished with material from the near-critical rotating star.
On the other hand, our data is inconsistent with wind compression
models predicting a strong outflowing velocity component.

\subsection{MULTI-LINE-TRANSITION INTERFEROMETRY: PHYSICAL CONSTRAINTS}
\label{sec:multitransition}

Obtaining spatially resolved information in multiple transitions of a
line series, such as the Balmer, Brackett, and Pfund series of 
hydrogen, opens fascinating new opportunities to determine physical
parameters, such as the excitation and ionization structure of the
line-emitting gas.

\begin{figure}[t]
  \centering
  \includegraphics[width=13.5cm]{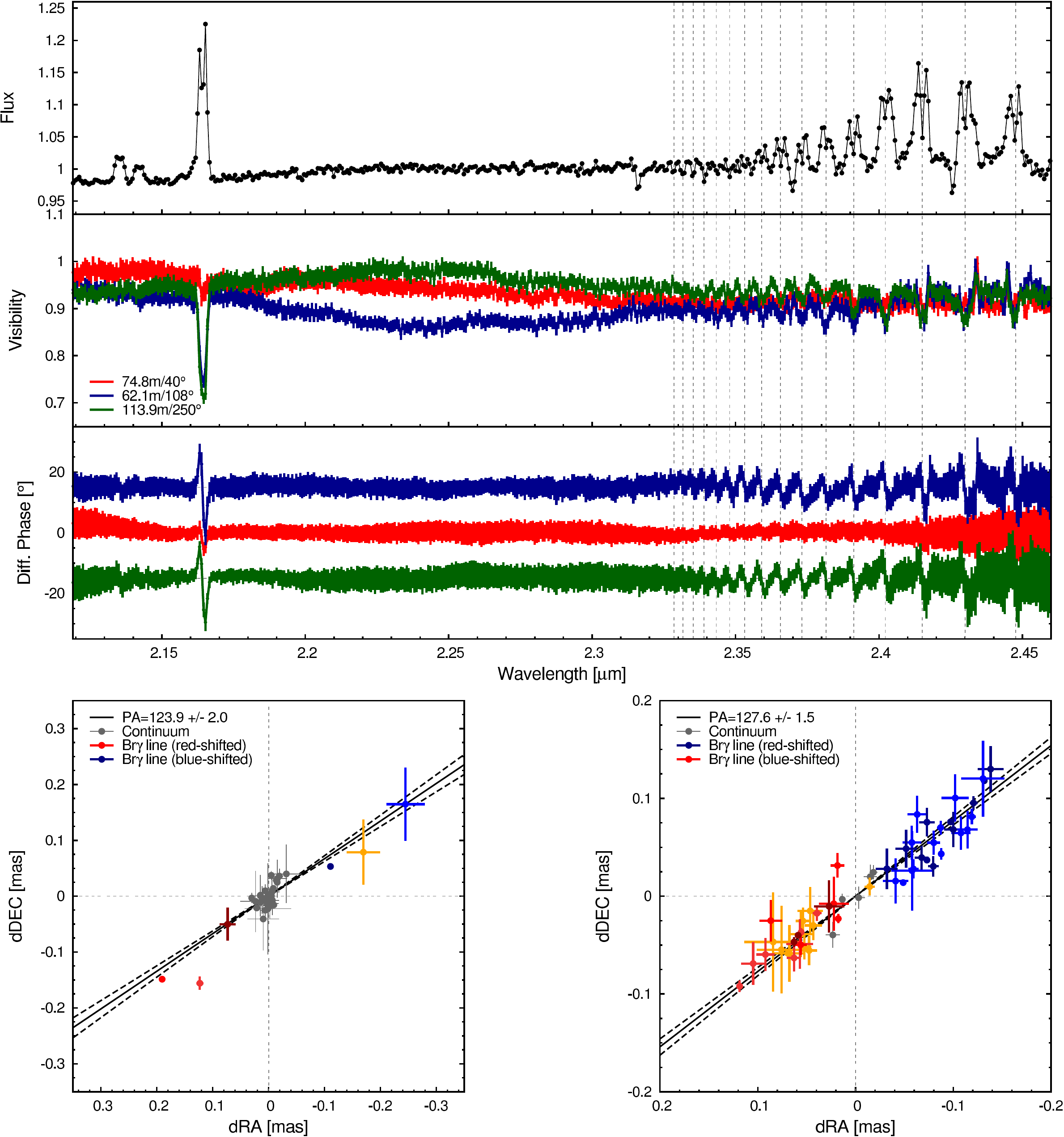}
  \caption{
    VLTI/AMBER MR-K2.3 observation on $\zeta$~Tau, covering with a single setup
    line transitions of Mg\,II 2.1374~$\mu$m, Mg\,II 2.1438~$\mu$m, 
    Br$\gamma$, and at least nine Pfund lines.
  }
  \label{fig:ZTau1}
\end{figure}

We applied this approach to the classical Be star
$\zeta$~Tau, which has already been extensively studied
with spectro-interferometry in the {H$\alpha$\cite{qui94,qui97,vak98,tyc04}}
and {Br$\gamma$\cite{ste09,car09}} recombination line.
Our new {observations\cite{kra12a}} were conducted using the
VLTI/AMBER MR-K2.3~mode ($R=1500$) and cover now, for the first time,
the Pfund emission lines associated with this object.
In the data, we detect at least nine line transitions from the Pfund
line (Pf14~2.4477~$\mu$m to Pf22~2.3591~$\mu$m) as well as the
Br$\gamma$ 2.166~$\mu$m line. 
Higher Pfund transitions are also present in the
spectrum, but cannot be clearly separated due to the broad,
double-peaked profile of the individual lines. 
Each double-peaked line clearly shows an S-shaped DP signature,
allowing us to derive the disk rotation axis for each line transition
separately, enabling tighter constraints on the disk position angle
(Fig.~\ref{fig:ZTau1}, bottom right) than from Br$\gamma$ alone
(Fig.~\ref{fig:ZTau1}, bottom left). 

\begin{figure}[t]
  \centering
  \includegraphics[width=8cm]{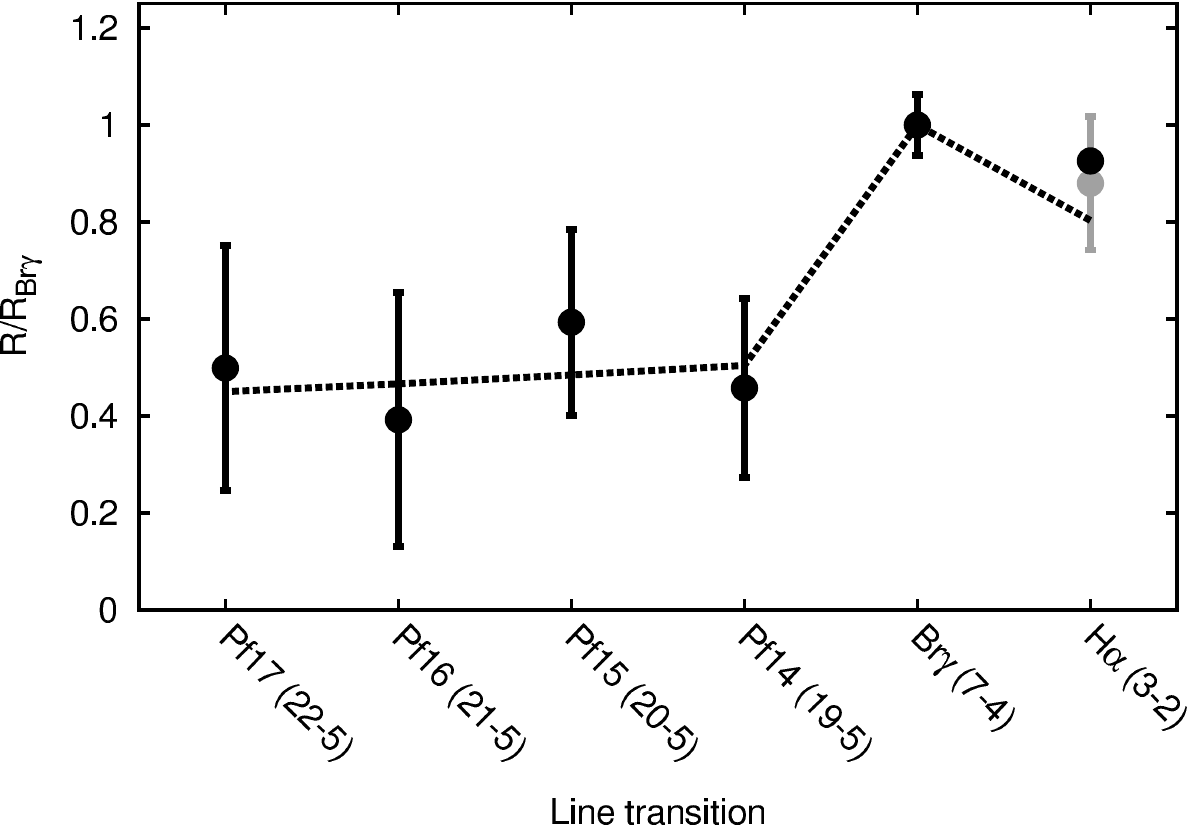} \hspace{1cm}
  \includegraphics[width=7.5cm]{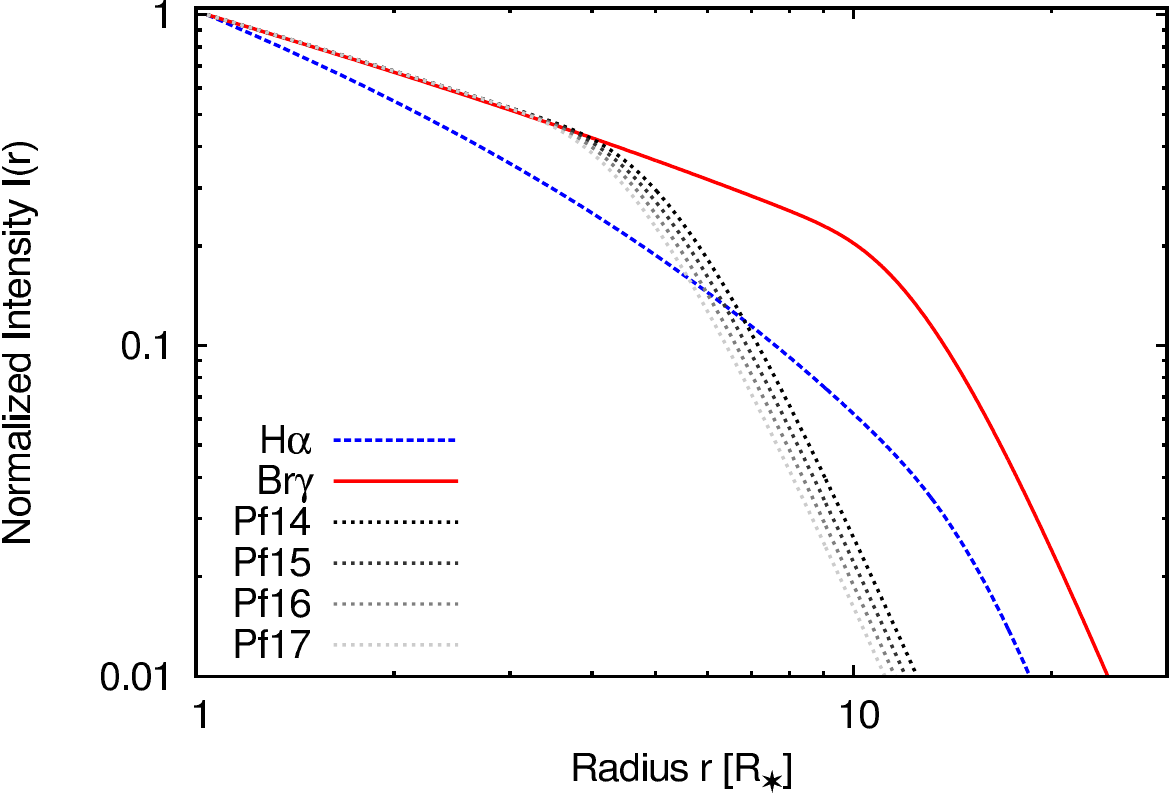}
  \caption{
    {\bf Left:} Spectro-interferometric measurements
    obtained for $\zeta$~Tau in multiple line transitions of the
    {Balmer\cite{qui94,qui97,vak98,tyc04}}, {Brackett\cite{kra12a}}, and
    Pfund {series\cite{kra12a}}.  The data points show the
    stellocentric emission radius, normalized to Br$\gamma$, while the
    dashed line shows the predictions from our LTE radiative transfer
    computation. 
    {\bf Right:} Radial intensity profile computed for $\zeta$~Tau
    using our LTE radiative transfer model.
  }
  \label{fig:ZTau2}
\end{figure}

In order to constrain the disk ionization structure, we derive the
spectro-interferometric emission radius for each line transition from
the continuum-corrected photocenter displacement vectors and then
compare them with the intensity profile derived from our LTE radiative
transfer {model\cite{kra12a}}.
For this modeling, we also include spectro-interferometric
measurements in the {H$\alpha$-line\cite{qui94,qui97,vak98,tyc04}},
which reveals that the H$\alpha$ and Br$\gamma$-lines trace
similar stellocentric emission radii, while the Pfund emission 
originates closer to the stars (Fig.~\ref{fig:ZTau2}, left).
This trend is reproduced by our LTE model (Fig.~\ref{fig:ZTau2},
right).
Our results confirm the finding from Pott et {al.\cite{pot10}}, which 
found a similar trend for the classical Be star 48\,Lib, and 
suggested that the measured size differences can be explained with 
optical depth differences between these line transitions.

Future studies might extend this approach to additional line
transitions and line tracers in other wavelength bands.
Together with sophisticated radiative transfer simulations, such
observations might reveal the excitation structure of the disk and
constrain parameters such as the temperature profile and the 
vertical disk structure, which are difficult to access with other
techniques.

\section{COMBINED SPECTRO-INTERFEROMETRY AND SPECTRO-ASTROMETRY}
\label{sec:combined}

Spectro-interferometry can provide a full characterization of the sub-AU
gas distribution and kinematics, but is also rather demanding in
telescope time and poses strict technical requirements, for instance
on the performance of the phase tracker.
As a result, most current spectro-interferometric studies still suffer
from a poor $uv$-coverage or compromise on the spectral dispersion.
In order to optimize the scientific constraints under these
conditions, it seems promising to combine spectro-interferometry with
spectro-astrometry, which is another technique that provides 
kinematical information about the gas distribution on AU scales.
Spectro-astrometry uses high-SNR long-slit spectra to measure the centroid
position of an unresolved object as function of wavelength. Since the
centroid position can be measured with much higher precision than the
size of the point-spread {function\cite{bai98a}}, this method allows
one to measure photocenter displacements of less than one
milliarcsecond in spectrally resolved emission lines.

In a recent study on the B[e] star {V921\,Sco\cite{kra12b,kra12c}}, we
combined the two techniques, for the first time, in a consistent way
for quantitative modeling, making use of the fact that the photocenter
offsets measured with spectro-astrometry are mathematically
equivalent to the wavelength-differential phases (DPs) measured in
spectro-interferometry. 

Our VLTI/AMBER data set on V921\,Sco consists of two observations with
high spectral dispersion ($R=12,000$, telescope triplet UT2-UT3-UT4)
and 24 AMBER observations with low spectral dispersion ($R=35$) in the
$H$- and $K$-band.
The low spectral dispersion data covers projected baseline lengths
in the range 10...127\,m with a good $uv$-coverage, which allows us
to reconstruct aperture synthesis images for three wavelengths bands
(centered around 1.6/2.0/2.3~$\mu$m) using the Building Block
{Mapping\cite{hof93}} image reconstruction algorithm.

\begin{figure}[t]
  \centering
  \includegraphics[width=16cm]{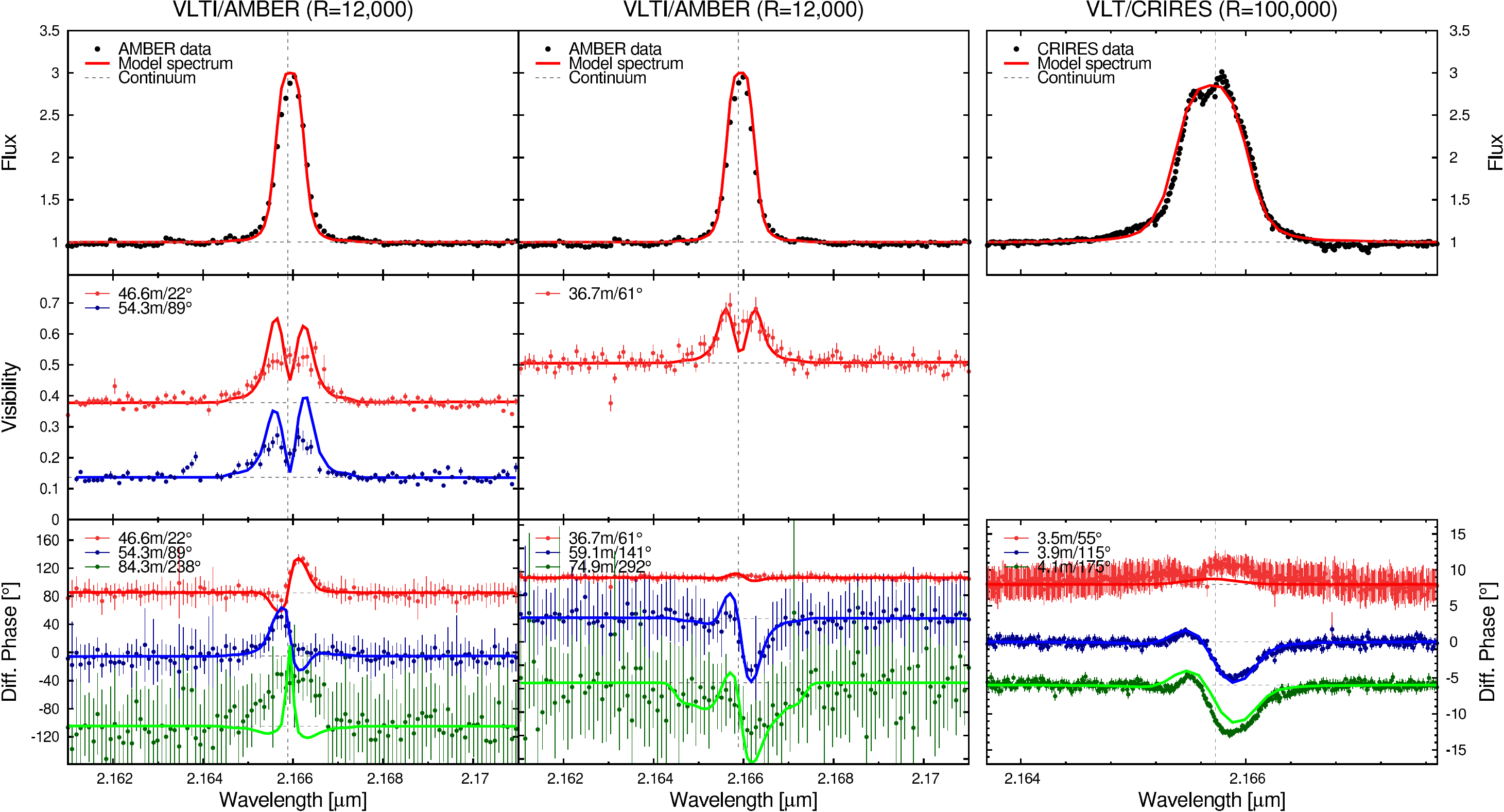}\\     
  \includegraphics[width=16cm]{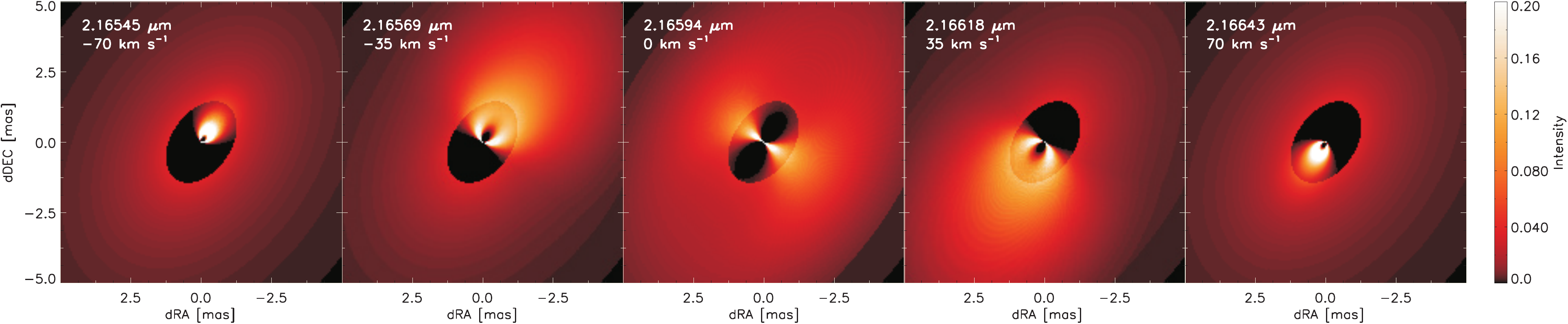}     
  \caption{
    {\bf Top:} Kinematical model for {V921\,Sco\cite{kra12c}},
    including AMBER spectro-interferometric (left and middle panel) and
    CRIRES spectro-astrometric data (right panel).
    {\bf Bottom:} Channel maps corresponding to our
    best-fit kinematical model, computed for five representative wavelengths.
    Besides the line emission, our model also includes the
    photospheric emission from the primary and secondary star and the
    thermal emission from the circumprimary dust disk. 
  }
  \label{fig:V921Sco-model}
\end{figure}

Complementary spectro-astrometric observations were obtained
with the AO-fed high-resolution infrared spectrograph
{CRIRES\cite{kae04}} at VLT. With a slit width of 0.2\arcsec, these
observations yield a spectral resolution $R=100,000$.
The spectra were recorded toward three different PAs ($55^{\circ}$,
$115^{\circ}$, $175^{\circ}$) and the corresponding anti-parallel PAs
($235^{\circ}$, $295^{\circ}$, $355^{\circ}$). In the data reduction
process, the spectro-astrometric signal derived from the parallel and
anti-parallel slit positions are subtracted in order to reduce
potential {artifacts\cite{bra06}} and the spectro-astrometric signal
is corrected for continuum {contributions\cite{pon08}}.
We then translate the measured spectro-astrometric signal into the
equivalent DP, which allows us to model the AMBER observables (line
profile, visibility, DP) and CRIRES observables (line profile, DP)
simultaneously using our kinematical modeling code.
Given its higher spectral resolution, the CRIRES data also reveals a
double-peak in the line profile (Fig.~\ref{fig:V921Sco-model}, right,
top panel; peak separation $\sim 0.000184~\mu$m or $\sim
25$~km~s$^{−1}$) that was not detected in the AMBER data.

\begin{figure}[t]
  \centering
  \includegraphics[width=16cm]{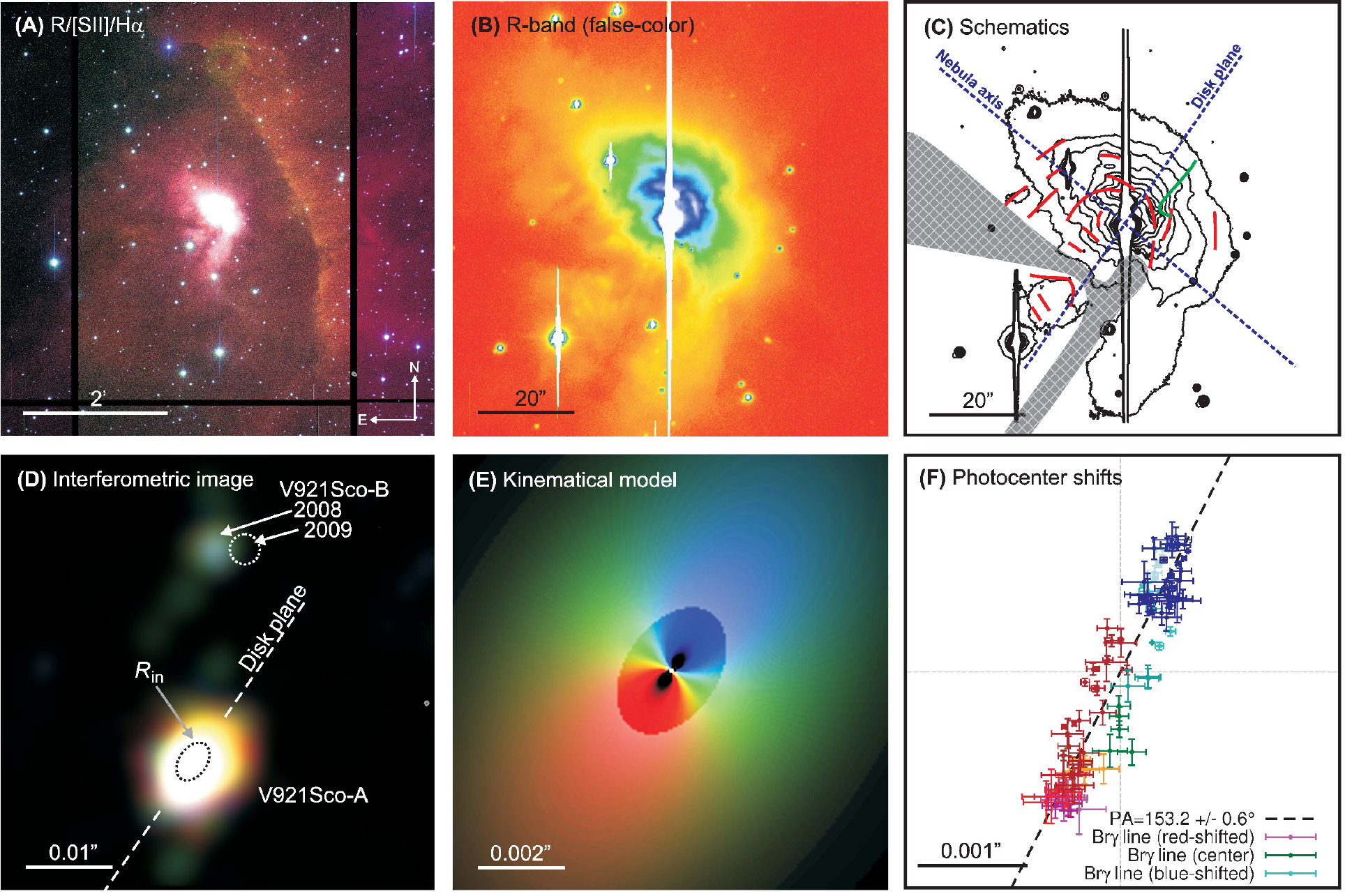}
  \caption{
    When combined with complementary techniques,
    spectro-interferometry enables an unprecedented global view of
    astrophysical systems, such as demonstrated here for V921\,Sco
    (see Sect.~\ref{sec:combined} for details).
  }
  \label{fig:V921Sco-overview}
\end{figure}

Besides the spectro-interferometric and spectro-astrometric data, we
obtained additional complementary data for V921\,Sco using the
Magellan/FIRE spectrograph and the Magellan/IMACS wide-field imager.  
These various interferometric and non-interferometric constraints
enable us to draw now a very comprehensive, global picture of the
V921\,Sco system, covering spatial scales from {300\arcsec} to
{0.001\arcsec} with a velocity resolution of up to 3~km\,s$^{-1}$. The
wide-field images show that V921~Sco is embedded 
in a bipolar nebula (Fig.~\ref{fig:V921Sco-overview}A), in which we
detect multi-layered, shell-like substructures
(Fig.~\ref{fig:V921Sco-overview}B+C) that might have been
shaped by episodic mass-loss events. Assuming the expansion speed 
derived from optical line profiles, we estimate the time scale between
different mass loss events to $\sim25...50$ years. 
Our VLTI/AMBER continuum aperture synthesis imaging reveals an
AU-scale disk-like structure that is oriented perpendicular to the
polar-axis of the arcminute-scale bipolar nebula (Fig.~\ref{fig:V921Sco-overview}C+D). 
Fitting analytic disk models, we find indications for a radial
temperature gradient and a central opacity depression, as expected for
an irradiated dust disk. Furthermore, we discover in the images a
close companion at a separation of $24.9\pm0.3$~mas 
(corresponding to $\sim 29$~AU at 1.15~kpc). 
Between two epochs in 2008 and 2009, we detect orbital motion of
$7^{\circ}$, implying an orbital period of $\sim 50$~years 
(assuming a circular orbit). Given the good agreement in the
time scales, we suggest that dynamical interaction between the newly
discovered companion and the circumprimary disk might trigger the
mass-loss events that shape the shell-like substructure in the ambient
bipolar nebula. This outflowing material might also provide the
required low-density conditions for the formation of the
B[e]-characteristic forbidden line emission. 

The high spectral dispersion VLTI/AMBER spectro-interferometric
and VLT/CRIRES spectro-astrometric observations allow us to  
measure the distribution and kinematics of the hot hydrogen gas in the 
system. From the derived photocenter shifts we derive the orientation
of the disk rotation plane and the spatial
displacement between the blue- and red-shifted Br$\gamma$ line wings 
(Fig.~\ref{fig:V921Sco-overview}F).
The center of light for the line spectral channels and for the
continuum channels are significantly displaced with respect to each
other (Fig.~\ref{fig:V921Sco-photocenter}A).  This offset reflects the
presence of the companion star, which shifts the center of light in
the continuum emission towards the secondary (North-West).
We compare the derived rotation plane of the gas disk
($153.2\pm0.6^{\circ}$) with the orientation of the
continuum-emitting disk ($145.0 \pm 8.4$).  The good agreement between
the PAs indicates a rotation-dominated velocity field. 

Using our kinematical modeling code, we quantify the rotation profile
and find that the line-emitting gas is, to high accuracy, in Keplerian
rotation. The model is able to reproduce the spectro-interferometric
and spectro-astrometric data as 
well as the measured double-peaked CRIRES line profile
(Fig.~\ref{fig:V921Sco-model}, top) and shows that the line-emitting
region extends from a few stellar radii to about twice the dust
sublimation radius. The channel maps corresponding to the best-fit
model are shown in Fig.~\ref{fig:V921Sco-model} (bottom).
Our measurement of the gas velocity field also provides important new
information about the evolutionary age of V921\,Sco, where earlier
studies have proposed either a young (Herbig~B[e]) or an evolved
(supergiant B[e]) nature.  We interpret our finding of a purely Keplerian
velocity field as strong evidence for the pre-main-sequence
nature of the object, since the decretion disks in post-main-sequence
B[e]-stars are believed to exhibit a significant outflowing velocity
component. 

Furthermore, our new information about the spatial distribution of
the line-emitting material opens interesting new pathways for the
physical interpretation of conventional spectroscopic observations,
which we conducted with the Magellan/FIRE spectrograph.
We obtained $J$+$H$+$K$-band spectra and derived
the line flux for 61 hydrogen recombination lines, including
transitions from the Paschen, Brackett, and Pfund series.
Given that the recombination physics of hydrogen is well known, one
can use these line decrements in order to derive physical parameters
such as the electron density $N_{e}$, although conventional
spectroscopic studies are typically not able to utilize this
information since they lack information about the volume over which
the line-emitting material is distributed.
From our AMBER observations we estimate the volume of the 
Br$\gamma$-emitting material and then solve for the electron density
($N_{e} = (2...6) \times 10^{19}$m$^{−3}$) assuming Case~B
recombination.

To summarize, in our V921\,Sco study AMBER spectro-interferometry and
CRIRES spectro-astrometry provide highly complementary information: 
Spectro-interferometry enables us to characterize the continuum
geometry, including a characterisation of the dust disk geometry and
the companion properties, and to spatially resolve the detailed
geometry and kinematics of the circumstellar gas with spectral
dispersion $R=12,000$.
Spectro-astrometry, on the other hand, adds valuable information at
even higher spectral dispersion ($R=100,000$), enabling a detection of
the double-peak line profile, and adds first-order kinematical
constraints in the spatially unresolved regime.

\subsection{VLTI/AMBER PHASE SIGN CALIBRATION PROCEDURE}
\label{sec:phasesign}

In order to achieve a meaningful astrophysical interpretation of the
measured DP and CP information, it is essential
to associate the phase sign with the on-sky orientation.  
In principle, such a phase sign calibration can be derived from 
first principles, based on very careful considerations of the beam
arrangement and propagation, the instrument design, and the
phase sign treatment in the data reduction and modeling software.

Our V921\,Sco {data\cite{kra12b,kra12c}} now allows us to perform this
calibration for VLTI/AMBER using a purely empirical approach. 
This data set is particularly well suited for this purpose,
given that the object consists of a well-characterized close binary
system with an unequal brightness ratio, where only one of the
components is associated with Br$\gamma$-line emission.
The presence of the companion offsets the photocenter between the
continuum and the line channels in a predictable way, from
which we can deduce the correct DP sign.

\begin{figure}[t]
  \centering
  \includegraphics[width=12cm]{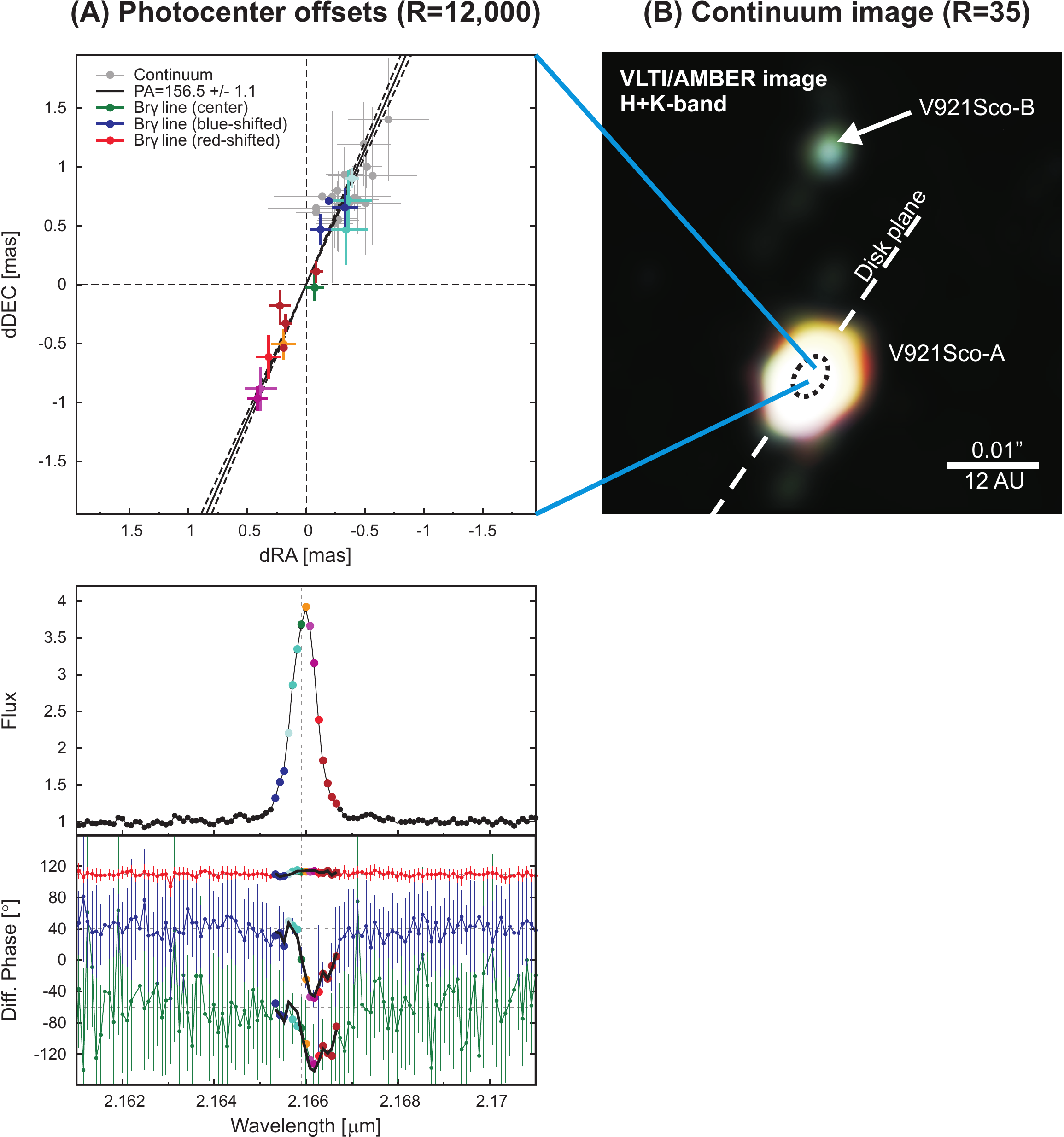}
  \caption{{\bf (A):} Spectro-interferometric observations on V921\,Sco in
    the Br$\gamma$-line with spectral dispersion $R=12,000$.
    The top panel shows the derived photocenter offsets, while the
    bottom panel shows the DP phases corresponding to the best-fit
    astrometric solution overplotted on the measurement.  In the
    photocenter plot, the origin of the coordinate system has been
    shifted to coincide with the center of light of the
    Br$\gamma$-line channels.
    {\bf (B):} Multi-color (R/G/B: 1.65/2.1/2.35~$\mu$m) aperture
    synthesis image reconstructed from low-spectral resolution data
    ($R=35$), revealing the presence of a close, previously unknown
    companion star.
  }
  \label{fig:V921Sco-photocenter}
\end{figure}

Specifically, we performed these steps:
First, we imaged the system using our $H$- and $K$-band AMBER
measurements with low spectral dispersion
(Fig.~\ref{fig:V921Sco-photocenter}B), which reveals that the
companion is located North-West of the primary star.  
This measurement does not suffer from any quadrant ambiguity,
given our earlier closure phase {calibration\cite{kra09a}} using the binary system
$\theta^1$\,Orionis~C.
From our AMBER high spectral dispersion measurements, we
determined the photocenter offset
and calibrated the DP sign based on the known quadrant of the companion
star (Fig.~\ref{fig:V921Sco-photocenter}A).

Using this new calibration, we determined the rotation senses of the
V921\,Sco circumprimary {disk\cite{kra12c}}
(Fig.~\ref{fig:V921Sco-photocenter}, left)
and of the {$\beta$~CMi\cite{kra12a}} disk (Fig.~\ref{fig:BCMi}).
Besides our high spectral dispersion observations ($R=12,000$) on
$\beta$~CMi, we observed $\beta$~CMi also in medium spectral
dispersion ($R=1500$), which allows us to extend our 
calibration also to AMBER's MR-mode, and apply it to our
$\zeta$~Tau data {set\cite{kra12a}} (Fig.~\ref{fig:ZTau1}), where we
find the opposite disk rotation sense as proposed by other
{workers\cite{ste09}}.

The proposed approach can be applied to calibrate any AMBER data
reduction+modeling code in a straight-forward manner and I provide
the data set for other AMBER users on the following website:
{http://www.stefan-kraus.com/files/amber.htm}

\section{SCIENCE-DRIVEN REQUIREMENTS FOR FUTURE INSTRUMENTATION}
\label{sec:requirements}

Spectro-interferometric beam combiner instruments with high
spectral dispersion (up to $R=12,000$ at near-infrared wavelengths
and $R=35,000$ at visual wavelengths) are in place since several years,
but could not yet exploit their full scientific potential.
Primarily, this is a result from the lack of phase tracking
instruments with sufficient sensitivity, preventing the study of
particularly interesting object classes, such as jet-driving T~Tauri
stars or Active Galactic Nuclei.
Therefore, throughput-optimized second-generation fringe tracking
instruments are urgently needed.  Promising strategies to optimize the
fringe tracking performance have already been proposed, including
off-axis tracking on a bright reference star (e.g. VLTI/PRIMA) or
fringe tracking and fringe recording in separate wavelength bands
(e.g. VLTI/MATISSE for the $L$+$M$+$N$-band and CHARA/VEGA $V$-band
interferometry with MIRC $H$-band or CLIMB $K$-band fringe tracking).

In Sect.~\ref{sec:photocenter}, I have outlined another severe
problem that arises from the fact that the visibility amplitudes in
spectro-interferometric observations are typically degraded by
residual phase jitter from the fringe tracker, preventing the 
measurement of well-calibrated visibility amplitudes. 
Future instrumentation projects should aim to eliminate this problem,
for instance by implementing a fast-switching mode between low and
high spectral dispersion.

Other desirable improvements would include an increased number of
combined apertures and longer baseline lengths,
which would enable a more effective sampling of the $uv$-coverage and
ultimately interferometric imaging in spectral lines.  
First attempts in this direction have been presented 
for the interacting binary {$\beta$~Lyrae\cite{sch09b}}
and the supergiant B[e]-star {HD\,62623\cite{mil11}}.
These studies used a self-calibration approach to recover the phase
information from NPOI and VLTI/AMBER data.
It is clear that much improvements in this field can be
expected in the future, both by enabling an improved uv-coverage,
but also from algorithmic work with dedicated spectro-interferometric
image reconstruction algorithms that could make full use
of the rich available DP information.

\section{CONCLUSIONS}
\label{sec:conclusions}

In this paper, I have outlined some of the unique capabilities
of spectro-interferometric instruments and discussed strategies for
the interpretation of such data, ranging from a photocenter and
position-velocity analysis approach to full kinematical modeling.
Particularly powerful constraints can be derived from
spectro-interferometric observations in multiple line transitions,
for instance in the hydrogen Balmer, Brackett, or Pfund series,
providing constraints on the ionization structure and electron density
of the line-emitting material. 

Furthermore, I have emphasized the complementary nature of 
spectro-interferometry and spectro-astrometry.
Spectro-astrometry provides a  resource-efficient method to 
measure first-order kinematical information with very high spectral
dispersion in the spatially unresolved regime.
Spectro-interferometry offers an even higher differential
astrometric accuracy (approaching the micro-arcsecond level)
and provides the ultimate technique to constrain the underlying
geometry of the line-emitting region through visibility amplitudes, a
task that cannot be achieved with any other technique in a
model-independent fashion.

\acknowledgments     

This work was performed in part under contract with the California Institute of Technology (Caltech) 
funded by NASA through the Sagan Fellowship Program.


\bibliography{spectrointerferometry}   
\bibliographystyle{spiebib}   

\end{document}